# Spin transport properties of spinel vanadate-based heterostructures

Antonio Peña Corredor[1], Alberto Anadón[2], Laurent Schlur[1], Jérôme Robert[1], Héloïse Damas[2], Juan-Carlos Rojas-Sánchez[2], Sébastien Petit-Watelot[2], Nathalie Viart[1], Daniele Preziosi[1]⋅, Christophe Lefevre[1]*

[1]Institut de Physique et chimie des Matériaux de Strasbourg, 67200 Strasbourg, France
[2]Institut Jean-Lamour, 54011 Nancy, France

*Corresponding author: christophe.lefevre@ipcms.unistra.fr
⋅Corresponding author: daniele.preziosi@ipcms.unistra.fr

Spin-orbit coupling and breaking of inversion symmetry are necessary ingredients to enable a pure spin current-based manipulation of the magnetization via the spin-orbit torque effect. Currently, magnetic insulator oxides with non-dissipative characteristics are being explored. When combined with non-magnetic heavy metals, known for their large spin-orbit coupling, they offer promising potential for energy-efficient spin-orbitronics applications. The intrinsic electronic correlations characterizing those strongly correlated oxides hold the promises to add extra control-knobs to the desired efficient spin-wave propagation and abrupt magnetization switching phenomena. Spinel vanadate $FeV_2O_4$ (FVO) exhibits several structural phase transitions which are accompanied by an intricate interplay of magnetic, charge and orbital orderings. When grown as a thin film onto $SrTiO_3$, the compressive strain state induces a perpendicular magnetic anisotropy, making FVO-based heterostructures desirable for spin-orbitronics applications. In this study, we have optimised the deposition of stoichiometric and epitaxial Pt/FVO heterostructures by Pulsed Laser Deposition and examined their spin-related phenomena. From angle-dependent magnetotransport measurements, we observed both Anisotropic Magnetoresistance (AMR) and Spin Hall Magnetoresistance (SMR) effects. Our findings show the SMR component as the primary contributor to the overall magnetoresistance, whose high value of 0.12% is only comparable to properly optimized oxide-based systems.

## 1. Introduction

Low-power electrical manipulation of magnetic states is one of the main goals of spintronics. Recent efforts have focused on ferromagnetic insulators (FMI) where a pure spin current can propagate without dissipation while exerting a torque on the magnetization itself[1,2]. A common scheme provides the use of the large spin-orbit coupling (SOC) in non-magnetic heavy metals (HM) grown adjacently to the insulating ferromagnet. A charge current flowing in the HM generates a transverse spin current via the spin Hall effect (SHE), while a spin accumulation generates a perpendicular charge current via the inverse spin Hall effect (ISHE). Thus, depending on the level of spin transparency at the HM/FMI interface, the magnitude of the spin accumulation, and the magnetic properties of the FMI, a spin-orbit torque (SOT) effect can take place and affect the FMI's magnetization (**M**)[3]. If **M** is perpendicular to the spin polarisation (**s**), the spin current will be absorbed and, if **M** is parallel to **s**, the spin current will be reflected and reinjected into the HM, thus impacting the current detected through the ISHE. Therefore, the HM's electrical resistance holds information of the relative orientation between **s** and **M** and varies with an applied magnetic field. This effect is known as Spin Hall Magnetoresistance (SMR), which serves as a proxy for spin-injection at the HM/FM interface as well as the detection of the magnetic state of the FMI. SMR plays a crucial role in identifying materials that have the potential for an efficient SOT switching, which is a fundamental requirement for the development of future energy-efficient and high-speed spin-orbitronic devices[4–6].

Robust SOTs-related phenomena are found in heterostructures that include materials with large SOC, large SHE and small spin diffusion length ($\lambda_{sd}$), such as the 5d metals[7]. As a result, in terms of material choice, platinum (Pt) is the most employed non-ferromagnetic HM for charge-to-spin conversion. To study this phenomenology with minimum proximity effects, the garnet ferrite $Y_3Fe_5O_{12}$ (YIG) is the most commonly chosen material, and Pt/YIG is the system where the SMR phenomenon was first experimentally reported[8]. Since then, several spinel ferrites ($Fe_3O_4$, $CoFe_2O_4$ or $NiFe_2O_4$) have also been chosen as FMI[9–12], and more recently, multiferroic compounds, such as $Ga_{0.6}Fe_{1.4}O_3$ and $Bi_{0.9}LaFeO_3$[13–15], have been investigated on the basis of a possible electric-field control of the magnetization, *i.e.* to challenge mainstream materials by adding extra functionalities in future FMI-based spin-orbitronic devices.

We introduce the spinel iron vanadate FeV$_2$O$_4$ (FVO) as a novel FMI material in the form of thin films. As bulk, FVO shows ferrimagnetism[16,17], ferroelectricity[18,19] (and hence a multiferroic behaviour[20,21]), an orbital-ordered state due to the presence of Jahn-Teller active cations (both Fe$^{2+}$ and V$^{3+}$) and a strong SOC[22–24]. More importantly for this study, FVO thin films exhibit a large perpendicular magnetic anisotropy (PMA) when grown onto STO single crystals as substrate. This mostly results from an enhanced tetragonality of the bulk cubic structure as caused by the STO-induced compressive strain and the strong magnetoelastic coupling, as shown in the case of FVO[26] and for other complex oxides thin films[27]. The effect of the compressive strain of the STO substrate modifies the crystal structure from a cubic spinel in bulk ($Fd\bar{3}m$) to a distorted tetragonal spinel ($I4_1/amd$) in the thin-film form. The paramagnetic-to-ferrimagnetic transition temperature is also modified by dimensionality effects and it is usually found at a larger value than the bulk (ca. 110 K)[25,28,29], as confirmed in this study (ca. 140 K - SM file).

In this work, we present an in-depth study of the magnetotransport properties of Hall-bar patterned Pt/FVO//STO heterostructures. SMR effects dominate the magnetoresistance signal, with a maximum value of 0.12% at low temperatures, which is of the same order of magnitude as the archetypical Pt/YIG system[30]. Links have been established between changes in the transport parameters and corresponding transformations within the FMI. The variation of the longitudinal and transverse resistivities have been explained via a competition of spin-driven events and other phenomena caused by magnetic anisotropy.

## 2. Experimental methods

Pulsed Laser Deposition (PLD) equipped with a KrF excimer laser (λ = 248 nm) was used to fabricate the Pt/FVO heterostructures onto SrTiO$_3$ (001) (STO) single-crystal substrates, and treated according to the standard Kawasaki procedure[31]. The FVO target was prepared by a standard solid state method[32,33]. FVO thin films were deposited at 400ºC, in an Ar pressure of $10^{-2}$ mbar, by ablating the target with a laser fluence of 4 J.cm$^{-2}$ at a repetition rate of 5 Hz, followed by a cooling down process in the same Ar partial pressure. For all heterostructures, the thickness of the epitaxial and stoichiometric FVO layer was of approximately 40 nm.

Pt was deposited at room temperature to avoid any inter-diffusion at the Pt/FVO interface, by ablating the Pt metallic target with a laser fluence of 4 J cm$^{-2}$, at a repetition rate of 5 Hz and in an Ar pressure of $10^{-3}$ mbar. The results presented here refer to samples with a Pt thickness of 3 nm, but similar results were observed for different thicknesses. The Pt layer was engineered into Hall-bars via a combination of UV-lithography and Ar-ion milling, as previously described[14]. Metallic top electrodes (10 nm Ti/150 nm Au) were evaporated to safeguard the Pt layer during the wire bonding process.

The surface quality of the Pt/FVO//STO heterostructures has been attested by atomic force microscope (AFM) measurements. A step-terraced profile characterized by a root mean square roughness ($R_q$) of ca. 0.2 nm, and reproducing that of the underneath STO substrate, was observed for all samples. The crystal structure of the heterostructures was determined by X-ray diffraction (XRD) experiments. Both the FVO and Pt layers thicknesses were verified by X-ray reflectivity measurements (Fig. S3 and Fig. S4 of the Supplementary material section (SM)). The in-plane (IP) and out-of-plane (OOP) lattice parameters for the FVO layer were found to be $a = b = 0.836$ nm and $c = 0.851$ nm at room temperature, respectively, thus indicating a substantial tetragonality if one considers the cubic FVO bulk value ($a_{bulk} = 0.8453$ nm). Both AFM and XRD measurements are shown in Fig. S1 and S2-3 of the SM.

The magnetic properties of the Pt/FVO//STO heterostructures were studied with a superconducting quantum interference device vibrating sample magnetometer (SQUID VSM MPMS 3, *Quantum Design*) after the Pt etching process. Their magnetotransport properties were studied with a cryo-free physical properties measurements system (Dynacool PPMS, *Quantum Design*). Magnetic fields up to 9 T were applied in different configurations, IP ($\alpha$) and OPP ($\beta$ and $\gamma$) [Fig. 1]. Measurements of both longitudinal ($V_{xx}$) and transverse ($V_{xy}$) voltages were performed.

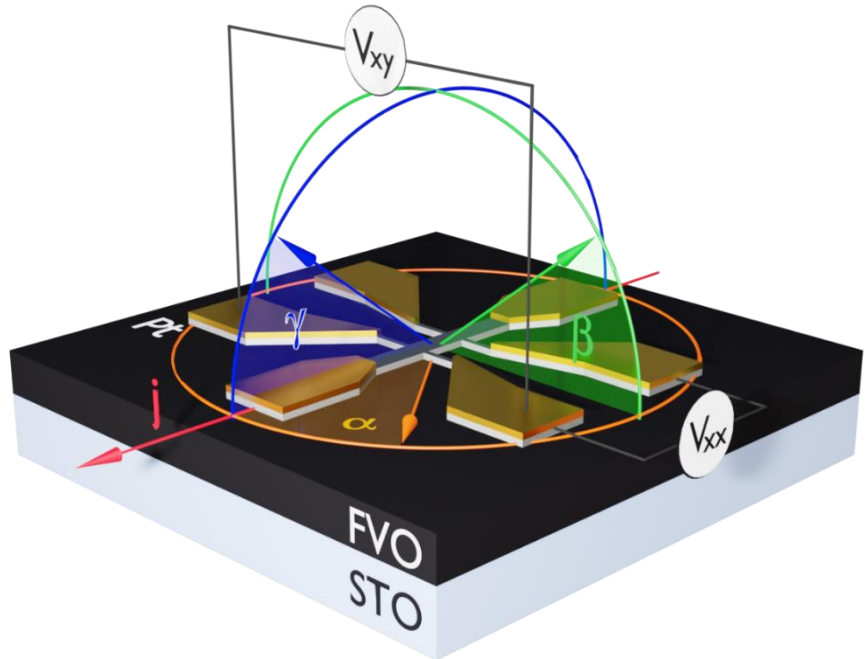

**Figure 1**. Representation of the Hall bar device, indicating the three different angular scanning configurations (α, β and γ) with the two different measured voltages ($V_{xx}$ and $V_{xy}$).

### 3. Results and discussion
### 3.1. Magnetic measurements

Figure 2a shows the magnetic hysteresis loop (MvsH) of our heterostructures, measured with a magnetic field applied in both in-plane IP and OOP directions at 10 K. The paramagnetic-to-ferrimagnetic transition was found to be around 140 K, in adequacy to previous reports on FVO//STO films[26,28] (Fig. S5 in the SM). The measured OOP MvsH loop is rectangular in shape, and thus, representative of a system showing a perpendicular magnetic anisotropy. Moreover, our OOP loop is characterized by two main features in line with already existing literature[26]: a step-like jump observed in the low field region and a large coercive ($H_c$) field. The small step-like jumps, observed in the OOP geometry, have been attribute to the presence of two magnetic phases with different coercivities coexisting within the same crystal structure, which seem to be strongly coupled in the IP direction but decoupled along the OOP [21]. The large $H_c$ is attributed to defects pinning the magnetic domains hindering their re-orientation [21]. Figure 2b shows the temperature dependence of $H_c$, as measured in OOP configuration. Their values rapidly drop upon increasing temperature and become zero for about 120 K. The magnetization at saturation ($M_s$) reaches 2 $\mu_B$ per formula unit (f.u.) at 10 K and is much larger in the OOP configuration than IP, in agreement with previous reports[26,28]. In a spin-only scenario with quenched orbital angular momentum, an antiparallel alignment of $Fe^{II}$ (4 $\mu_B$) and $V^{III}$ (2 $\mu_B$) would result in a net zero $M_s$. The non-zero value found is explained via a nonnegligible SOC in $V^{III}$, which reduces the cation's magnetic contribution to *ca.* 0.85 $\mu_B$ – measured in bulk FVO[16]. In our heterostructures, assuming a spin-only scenario for $Fe^{II}$ results in a net moment of around 1 $\mu_B$ per $V^{III}$ site.

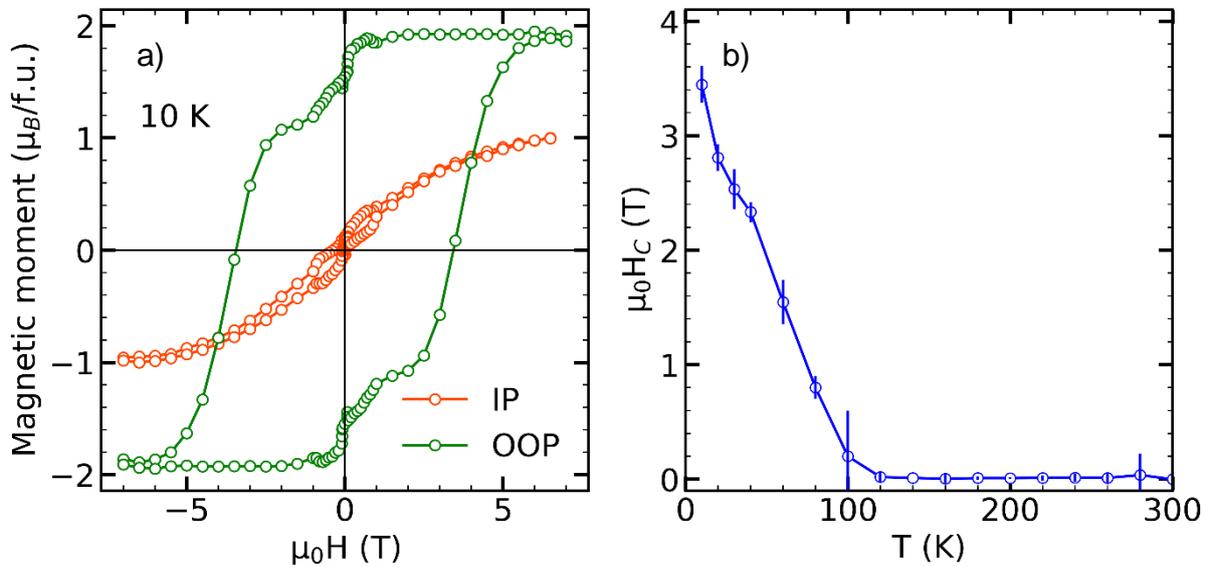

**Figure 2**. a) *M*vs*H* curves of the Pt(3 nm)/FVO(40 nm)//STO heterostructure in IP and OOP configurations at 10 K, showing strong PMA, b) Temperature evolution of the OOP coercive field, as extracted from OOP M vs. H curves.

### 3.2. Angle-dependent longitudinal magnetoresistance measurements

Figure 3a shows the temperature dependence of the longitudinal resistivity at zero magnetic field ($\rho_0$) for a Pt(3nm)FVO(40nm)//STO heterostructure. The resistivity presents no upturn at low temperature (down to 2 K), with a $\rho_0$ of ca. 3 $\times$ 10$^{-7}$ $\Omega$ m, while the $\rho_0$(300K) was found to be around five times the Pt bulk value ($\rho_b$), which is of the same order as measured for high-crystalline Pt films of the same thickness, and grown by magnetron sputtering ($\approx 3\rho_b$)[34]. This demonstrates the good growth optimization of the Pt layer onto FVO by PLD. Figure 3b displays the variation of the longitudinal resistance ($\rho_{xx}$) vs. magnetic fields applied in the OOP direction ($H_z$) at different temperature. In the 2 - 40 K temperature range, we observe a spin-valve-like effect most likely due to the two magnetic phases characterizing the FVO in the OOP direction (*cf.* Fig. 2a), with a negative magnetoresistance that saturates at high $H_z$ values. Around 120 K, and mostly depending upon the magnetic field value, the magnetoresistance becomes positive. This change can be clearly observed in Figure 3c, where the relative variation of the magnetoresistance $\Delta\rho_{xx}(H_z)/\rho_0 = (\rho_{xx}(H_z) - \rho_0)/\rho_0$ has been plotted vs. temperature. Such variation can be understood when considering a competition between the different magnetoresistance phenomena.

As discussed in Ref.[14], two main physical phenomena contribute to the magnetoresistance in the NM layer: the spin Hall magnetoresistance (SMR) and the anisotropic magnetoresistance (AMR). The SMR effect is based on the interaction between the generated spin current and the FMI moments, while the AMR depends on the magnetic proximity effect (MPE) exerted by the FVO magnetization on the Pt layer [14] and is therefore expected to shrink when temperature increases. The longitudinal resistivity as a function of the magnetization orientation of the FVO can be written as:

$$\rho_{xx} = \rho_0 + \Delta\rho_{AMR} m_x^2 + \Delta\rho_{SMR\|} m_y^2 \qquad (1)$$

where $\Delta\rho_{AMR}$ and $\Delta\rho_{SMR\|}$ are the AMR and parallel SMR contributions to the longitudinal resistivity and $m_x$ ($m_y$) is the component of the reduced magnetization in the x (y) direction (cf. Fig. 1).

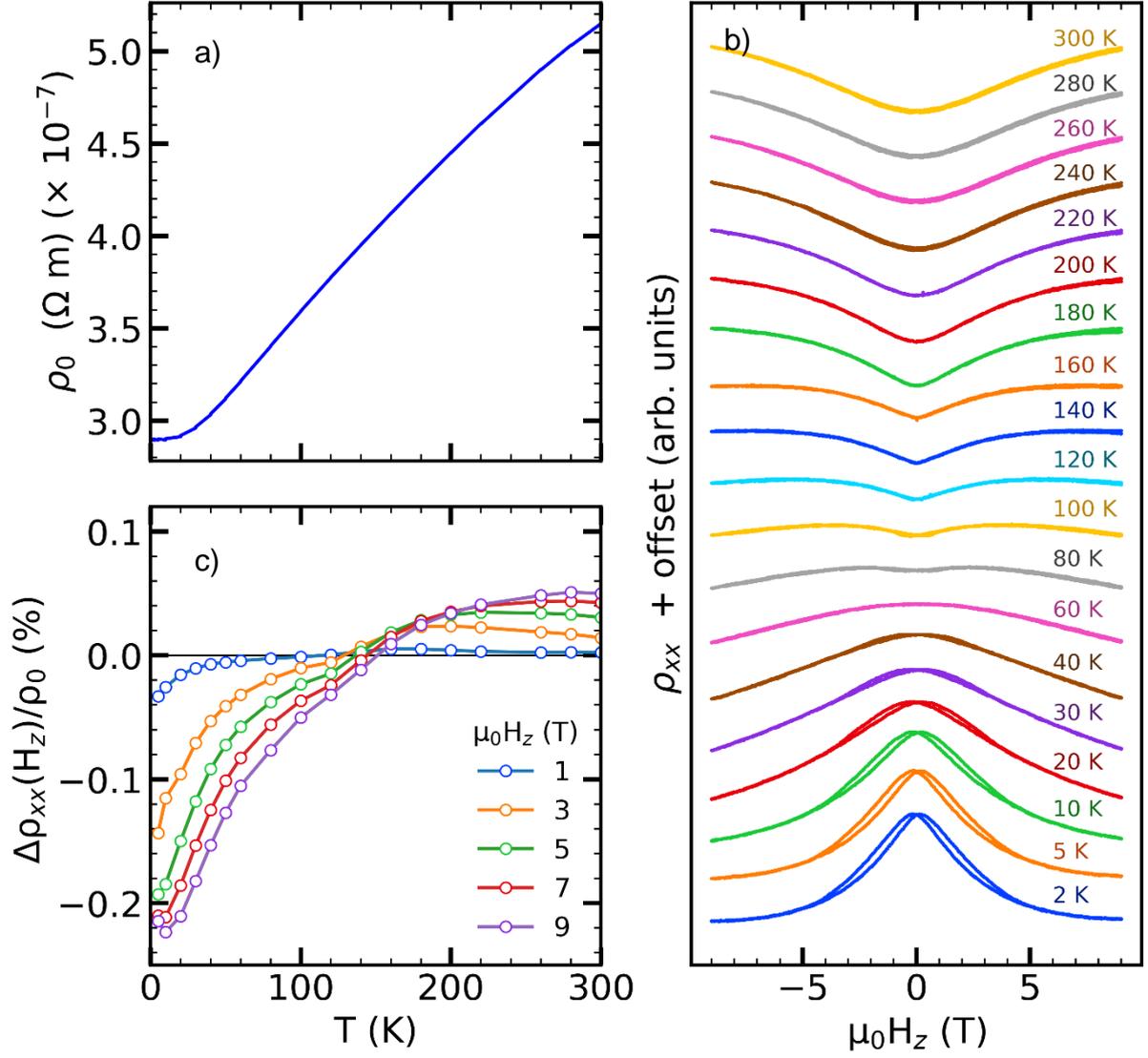

**Figure 3**. a) $\rho_{xx}$ vs. T. b) $\rho_{xx}$ as a function of $H_z$, for temperatures in the 2 - 300 K range. c) Calculated temperature dependence of the longitudinal magnetoresistance for different $H_z$. One can notice a sign change around 120 K.

According to **Eq.1**, by doing an appropriate angular choice, one can isolate the AMR from the SMR components. While the α configuration (*cf.* to Fig. 1), encompasses both SMR and AMR components, the β and γ ones only result from the SMR and AMR components, respectively.

Figures 4a, b, c show the angular dependences of the magnetoresistance $\Delta\rho_{xx}/\rho_0 = (\rho_{xx} - \rho_0)/\rho_0$ in the $\alpha$, $\beta$, and $\gamma$ configurations, with corresponding fits with a $A\cos^2(\theta + b)$ function, where $A$ represents the magnetoresistance's amplitude and $b$

the angle offset due to sample mounting. The azimuthal α scan, which simultaneously probes SMR and AMR components, does correspond with the addition of the β and γ scans – considering the angular phase difference.

In the plot of the scan amplitudes as a function of the temperature one can also observe that the IP rotation captures both variations. In the SM (Fig. S6) we show the equivalence between γ-scan magnetoresistance and $H_x$-scans. The AMR has a negative contribution to the magnetoresistance, whereas the SMR has a positive one. This confirms the opposing nature of the two phenomena, as inferred by the sign change of the $\Delta\rho_{xx}/\rho_0$ curves. The SMR amplitude is particularly large below the paramagnetic-to-ferrimagnetic transition temperature, but it keeps a non-zero value also above it. On the contrary, the AMR amplitude sharply decreases when temperature increases, until zeroing out at the paramagnetic-to-ferrimagnetic transition. The SMR is significantly larger than the AMR at all temperatures, showing that the spin-Hall-based component dominates the magnetoresistance response of our heterostructures. A maximum SMR value of 0.12% for a Pt/FVO sample has been obtained at 2 K, matching the maximum values obtained for properly optimized Pt/YIG samples.

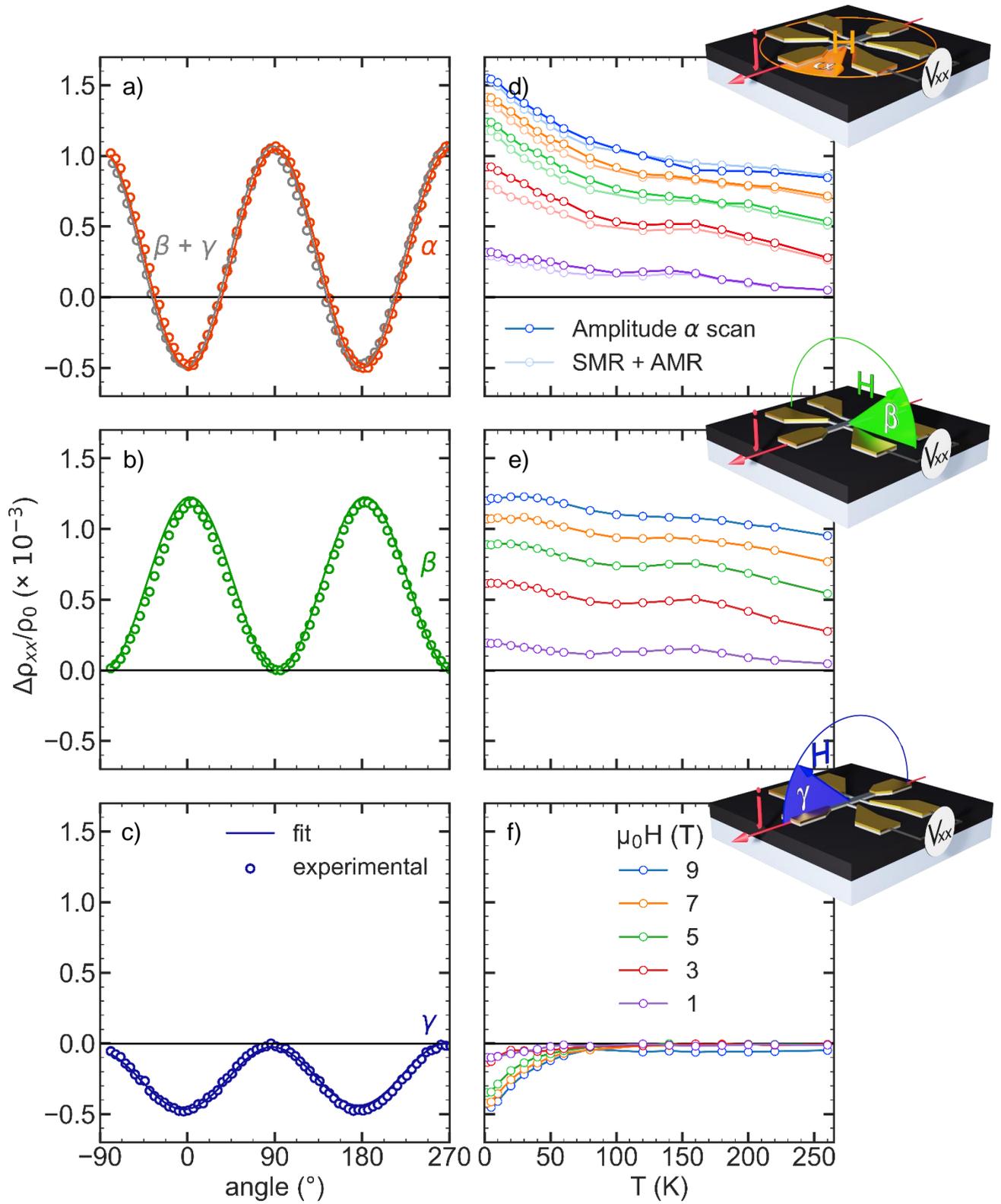

**Figure 4.** a), b) and c) angle-dependent measurements (α, β and γ) at 9 T and 2 K, solid lines are fit to the experimental data. d), e) and f) Amplitudes of the angle-dependent measurements as a function of temperature, for different magnetic field values.

**3.2. Angle-dependent transversal magnetoresistance measurements**

The transverse resistivity is the sum of some IP ($\rho_1$) and OOP ($\rho_2$) contributions, according to the following formula [30]:

$$\begin{aligned}\rho_{xy} &= \Delta\rho_1 m_x m_y + \Delta\rho_2 m_z = \\ &= (\Delta\rho_{PHE} - \Delta\rho_{SMR||})m_x m_y + (\Delta\rho_{SMR\perp} + \Delta\rho_{OHE} + \Delta\rho_{AHE})m_z\end{aligned} \quad (2)$$

The IP contribution ($\Delta\rho_1 = \Delta\rho_{PHE} - \Delta\rho_{SMR||}$), depends on the planar Hall effect (PHE) and on the parallel component of SMR (SMR||). The first phenomenon stems from the spin-orbit interaction in conducting ferromagnets, similarly to the AMR, while the latter is an interface spin-driven effect. The variation of $\Delta\rho_1$ with temperature been discussed in the SM (Fig. S7). The OOP contribution ($\Delta\rho_2 = \Delta\rho_{SMR\perp} + \Delta\rho_{OHE} + \Delta\rho_{AHE}$), depends on the perpendicular component of the SMR (SMR$_\perp$), the ordinary Hall Effect (OHE), and the anomalous Hall effect (AHE)[35,36].

Hall hysteresis loops (V$_{xy}$ *vs.* H$_z$) have been performed with $H_z$ in the − 9 to 9 T range and with temperature varying between 2 and 300 K, in order to characterise the $\Delta\rho_2$ component – depending on $m_z$. The experimental data have been corrected from the contribution of the OHE, and the resulting $\rho_2' = \Delta\rho_2 - \Delta\rho_{OHE}$ are odd with respect to H$_z$, as shown in Figure 5a. $\rho_2'$changes sign around 140 K (Figure 5b), coinciding with the magnetic transition in the material. The remanent resistivity at zero field ($\rho_{2R}'$), decreases with temperature, cancelling out as the ferrimagnet's magnetic order fades away. As illustrated in Figure 5c, the coercive field for the resistivity reversal (H$_{c-i}$),defined as the magnetic field at which $\rho_2' = 0$, increases with the temperature as the $\rho_2'$ hysteresis close up, and drops sharply to zero once ferrimagnetism disappears.

The OHE component was used to calculate the Hall constant (R$_H$), the charge mobility (μ$_e$) and the carrier density (n), as summarized in Figures 5 c,d. We observe that R$_H$ is negative for the entire temperature range, pointing at a conduction principally associated to electrons which density n drops at low temperatures, concomitantly at an increase of their mobility.

The sign-reversal for $\rho_2'$ vs. T has already been reported in the case of Pt/YIG [37]. On a phenomenological basis this has been attributed to a change in the imaginary part of the spin mixing conductance ($G_i$) [37], as described in the following formula [3] :

$$\frac{\rho_2'}{\rho_0} \approx \frac{2\lambda_{sd}^2 \theta_{SH}^2}{t} \frac{\sigma G_i \tanh^2 \frac{t}{2\lambda_{sd}}}{\left(\sigma + 2\lambda_{sd} G_r + \coth \frac{t}{\lambda_{sd}}\right)^2} \tag{3}$$

where $G_r$ is the real part of the spin mixing conductance (indicator on the spin scattering and transmission at the Pt/FVO interface), while $t$, $\sigma$ and $\theta_{SH}$ represent the Pt thickness, conductivity and spin Hall angle, respectively.

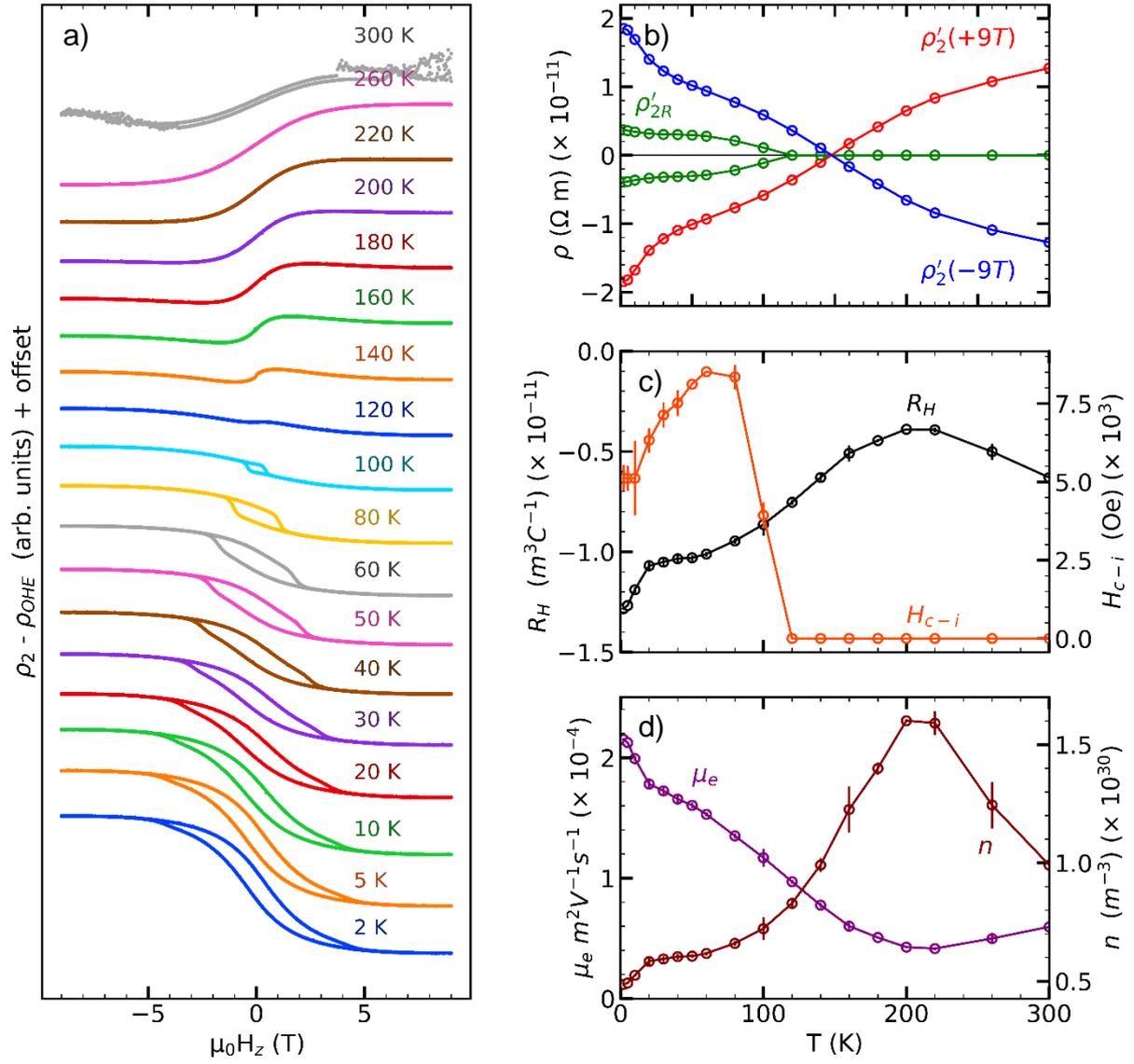

**Figure 5**. a) $\rho_2' = \Delta\rho_2 - \Delta\rho_{OHE}$ hysteresis loops at various temperatures with an artificial for better visualization. b) $\rho_2'$ at − 9 and 9 T vs. of T, with $\rho_2'_R$, the remnant resistivity at zero field. c) Hall constant value vs. T, and coercive field for $\rho_2'$ curves. d) Electron mobility and charge carrier density vs. T.

Differently from our system, in Pt/YIG the change in $\rho_2'$ sign occurs from positive at low temperature to negative at higher temperatures, with an inversion at around 100 K[37,38]. A similar trend has been reported by the authors for previously investigated Pt/Ga$_{0.6}$Fe$_{1.4}$O$_3$ heterostructures[14].

**Eq. (2)** indicates that the physical origin of this inversion is a competition between the $\Delta\rho_{SMR\perp}$ component, which depends on the nature of our Pt/FVO heterostructure, and the $\Delta\rho_{AHE}$ contribution, whose strength varies as the AMR, and is therefore affected by the MPE. The sign reversal for the transverse resistivity signal could be explained by assuming that the contribution of each mechanism is opposite to each other[9]. The MPE is known to be stronger at low temperatures, due to a strong exchange coupling between the FVO cations 3d and the Pt 5d orbitals[39], which weakens at higher temperatures. As a result, and at a certain threshold temperature, both $\Delta\rho_{SMR\perp}$ and $\Delta\rho_{AHE}$ contributions cancel each other, and beyond that point the former dominates in the transverse resistivity. Another explanation for that sign inversion, and discussed in Ref. [40] could originate in the paramagnetic state of the Pt layer for which both the curvature of the Fermi surface and the density of states drastically change with temperature.

## 4. Conclusions

The presented results constitute the first study of spin-Hall properties in Pt/FeV$_2$O$_4$//SrTiO$_3$ heterostructures. The system's large PMA is ideal for spin-orbitronic applications. Angle-dependent magnetotransport longitudinal measurements were performed to determine the AMR and SMR contributions to magnetoresistance, showing that the spin-based component dominates in all cases and reaches a value of 0.12% at 2 K, comparable to that obtained for largely optimized Pt/YIG heterostructures. This suggests efficient spin injection at the Pt/FVO interface, indicating the potential of FVO-based structures for SOT-induced magnetization switching. Transverse measurements have been conducted to quantify the anomalous component of the heterostructure. Its evolution with temperature can be associated with changes in FVO's magnetic behaviour, proving the influence of the ferrimagnet's magnetization onto the non-magnetic heavy metal above.

This work provides a first comprehensive study of the spin-dependent magnetotransport properties of Pt/FVO heterostructures, highlighting the potential of FVO as a promising candidate for future multifunctional spintronic devices.

# Supplementary material

## 1. AFM images

The surface of the STO substrate has been imaged by AFM (**Figure S1a**) after the treatment with HF using Kawasaki method[31] and high-temperature annealing (900ºC, 2h), which lead to a surface roughness of ca. 0.16 nm showing a step-terrace surface. The deposition of an FVO layer of ca. 30 nm resulted in a roughness of ca. 0.19 nm (**Figure S1b**) and the deposition of the Pt/FVO layers (**Figure S1c**) gave a roughness value of ca. 0.21 nm. The step-terrace structure is kept after both depositions. Roughness values have been obtained through surface region analyses of the surface areas shown in **Figure S1**.

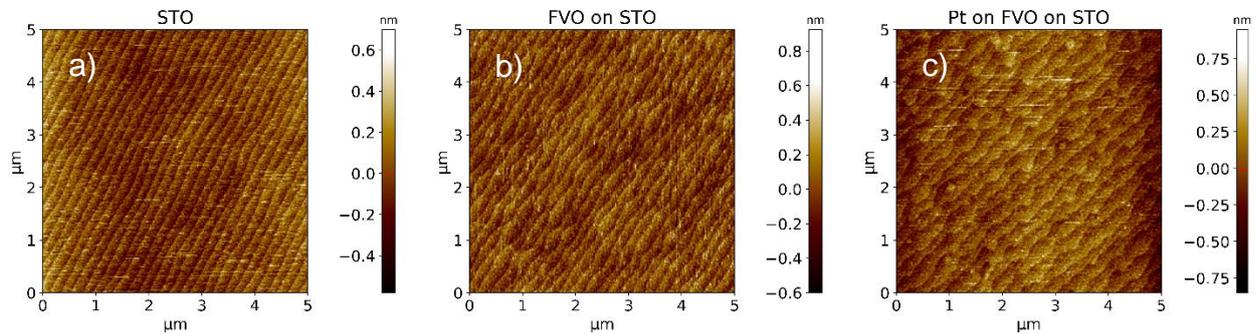

**Figure S1**. AFM images of the STO substrate after the HF treatment – a), and subsequent annealing, as well as that of the sample post FVO – b), and post Pt – c) depositions.

## 2. XRD measurements

The lattice parameters of the FVO layer have been determined by performing Reciprocal Space Maps (RSM) around the {448} reflection family (**Figure S2**), leading to the values indicated in the main text: $a = b = 0.836$ nm and $c = 0.851$.

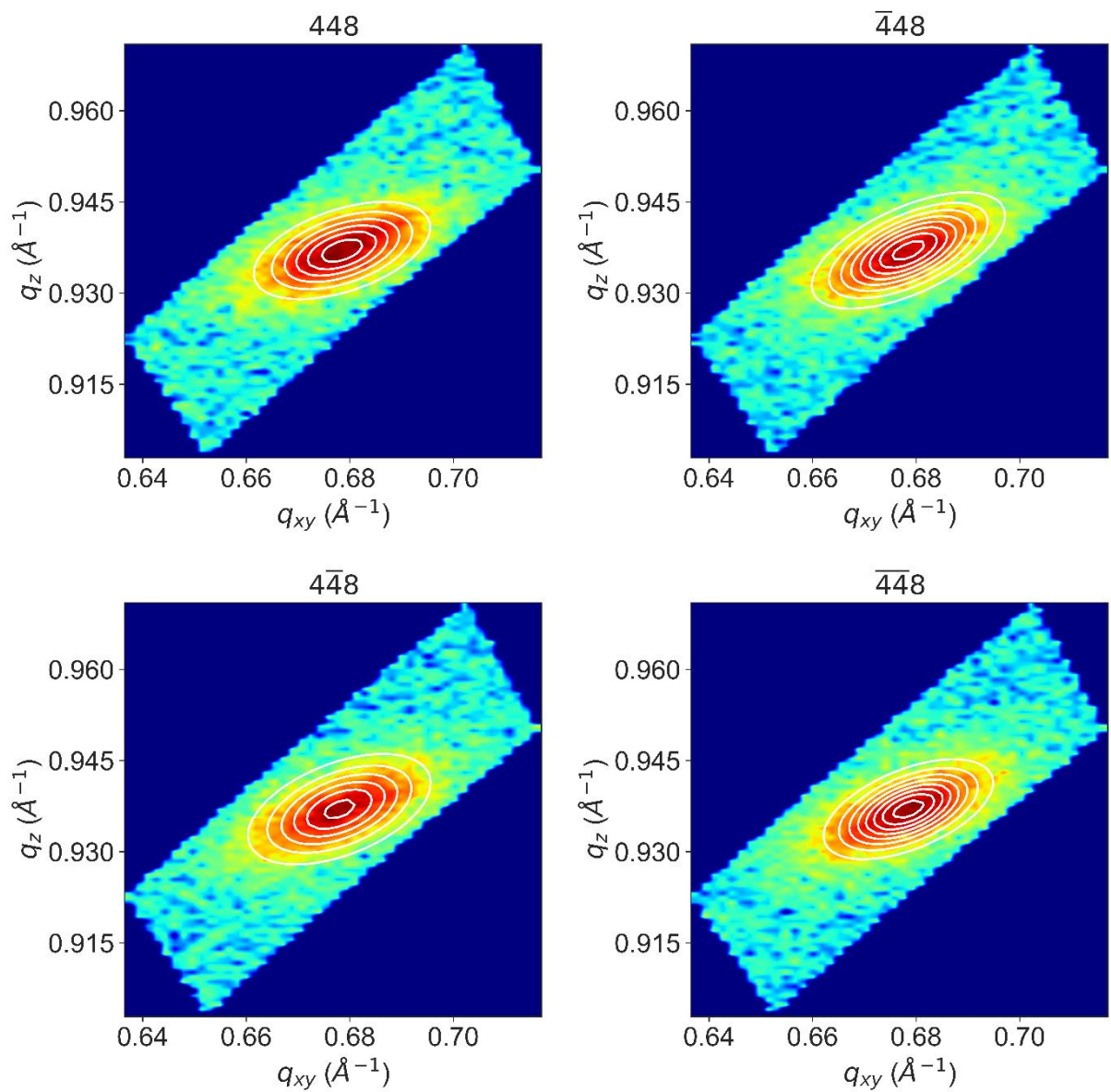

**Figure S2**. 2D-RSM images on the {448} reflections of FVO//STO films.

Theta-2theta measurements have been carried out to determine the out-of-plane lattice parameter on the Pt/FVO//STO heterostructures, as shown in **Figure S3**.

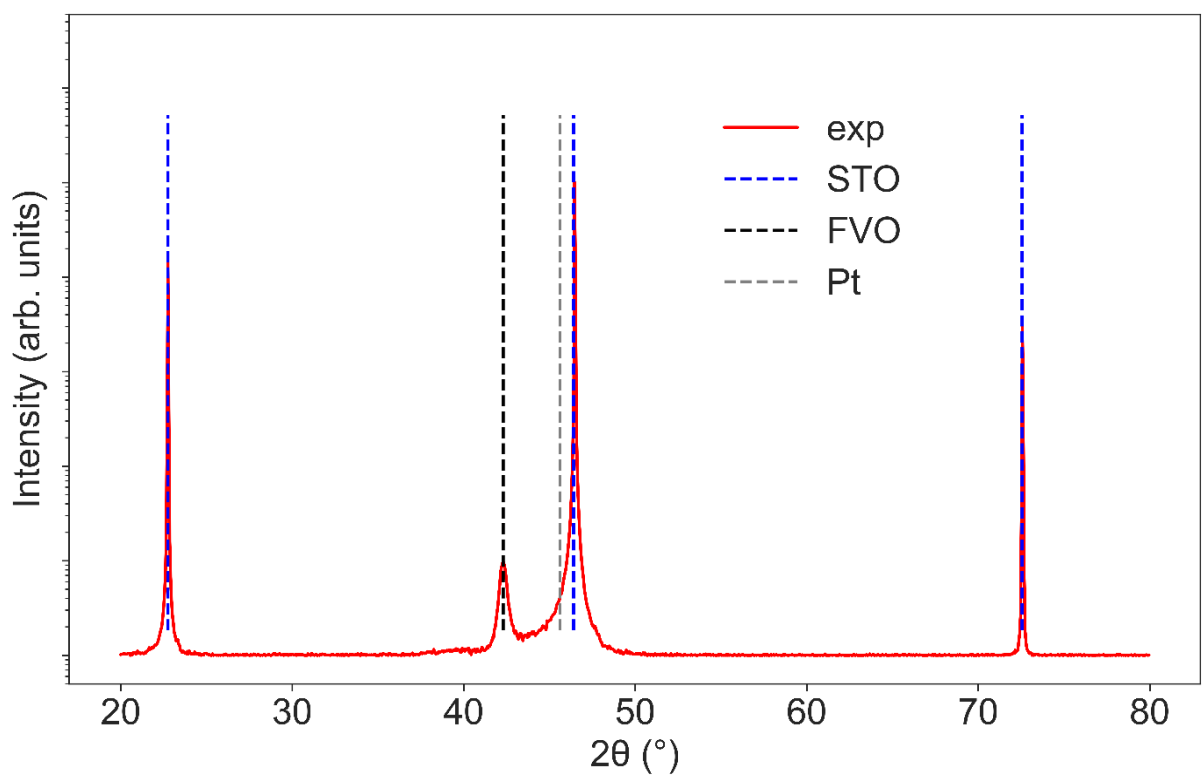

**Figure S3**. Theta-2theta diffractogram of the Pt/FVO//STO heterostructures, with the {00l} reflections.

The thickness of FVO in both FVO//STO thin films and Pt/FVO//STO heterostructures were determined by reflectivity measurements: **Figure S4** and **Figure S5**.

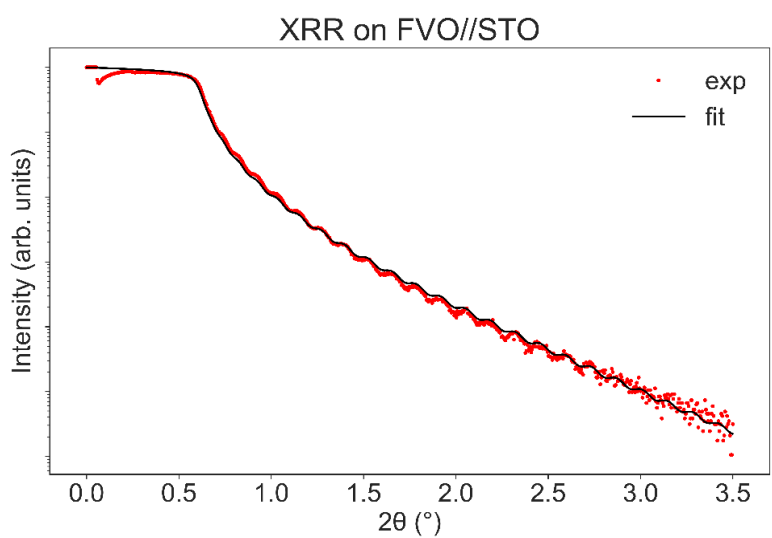

**Figure S4**. XRR – experimental data and best fit for a FVO//STO film. Fitted thickness: 39.1(7) nm.

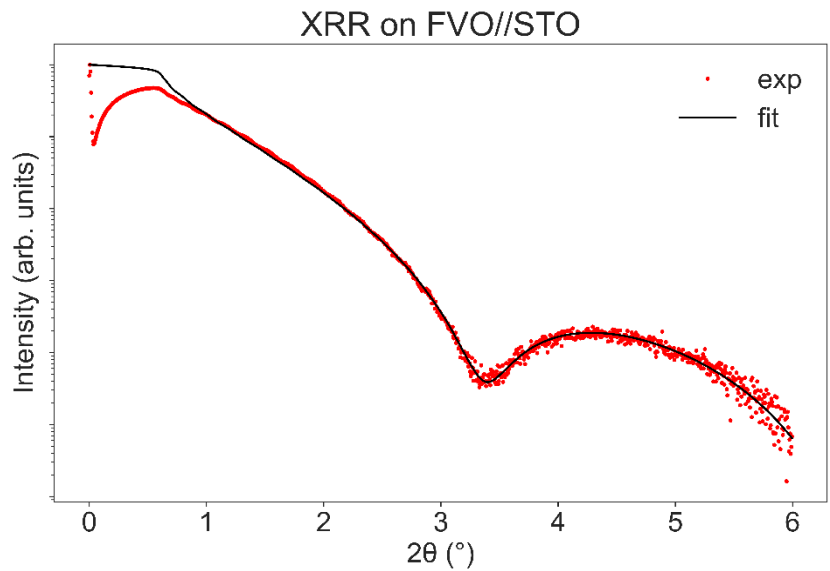

**Figure S5**. XRR – experimental data and best fit for a Pt/FVO//STO heterostructure. Fitted thickness for the Pt layer: 3.12(2) nm.

3. **Magnetization vs. temperature measurements**

Field-cooled (0.5 T) magnetization vs. temperature measurements have been conducted in order to determine the magnetization behavior as a function of temperature of the Pt/FVO//STO heterostructures. As shown in **Figure S5**, the magnetization profile indicates a ferrimagnetic-to-paramagnetic transition at around 140 K, coinciding with previous results in the literature[28,38] and the expected transition for bulk FVO[16,23]. However, as it can be seen in the raw data, the magnetization does not fall down to zero and a ferrimagnetic component might survive beyond the transition temperature with no coercivity (cf. Fig. 2).

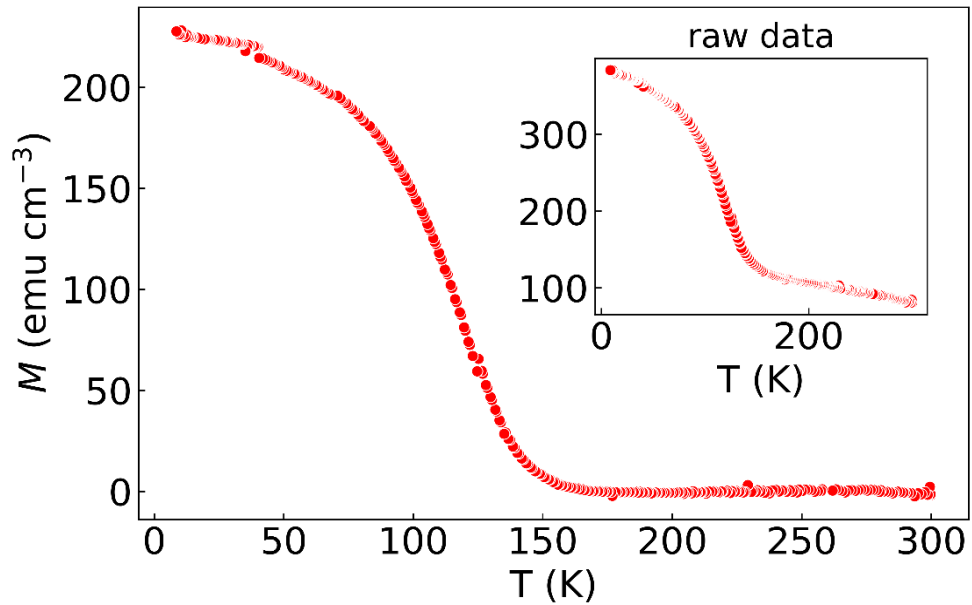

**Figure S5**. Magnetization vs. temperature for the Pt/FVO//STO heterostructure. Main plot: data with paramagnetic correction. Inner plot: raw data (as measured).

### 4. Comparison between angle-dependent measurements in the γ configuration and an Hx-scan for AMR

Angle-dependent measurements have been carried out in the **γ** configuration at 2 K at for 9 T, leading to the results shown in **Figure S6a.** At that same temperature, magnetic field scans have been conducted, as presented in **Figure S6b**. The amplitude for the oscillation matches the value of magnetoresistance for 9 T, which proves the equivalence between the two measurements.

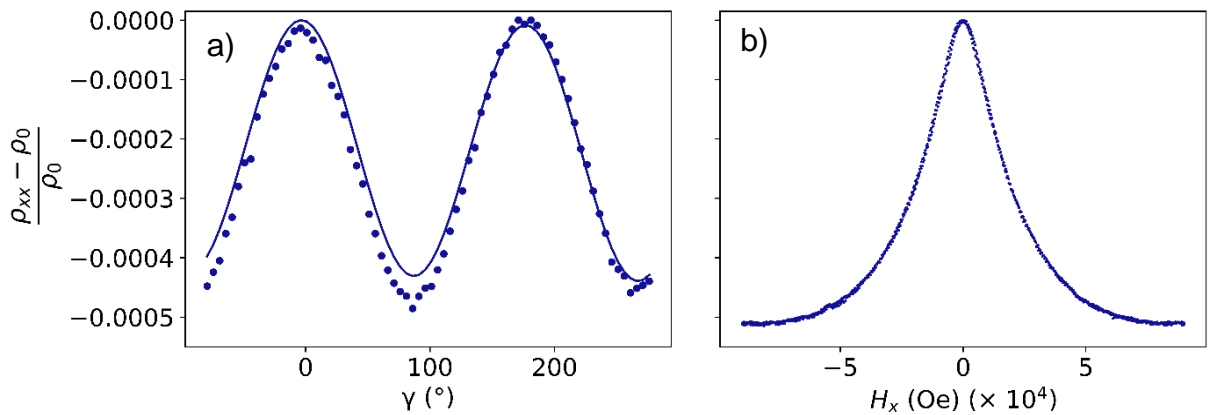

**Figure S6**. a) γ scan at 2 K and 9 T. b) H$_x$ scan at 2 K in the [-9 T, 9T] field range. The maximum amplitude coincides between the two measurements, which confirms the consistency of the analyses.

5. α **scans**

For the characterisation of the Δρ$_1$ component (as defined in Equation 2), in-plane ($\alpha$) scans were conducted with an applied field of 7 T for the characterisation of the ρ$_1$ component, as presented in **Figure S7**. The curves have been fitted using a $A \cos^2(\theta + b)$ function (**Figure S7a**) and the oscillation amplitude $A$ has been represented vs. T in **Figure S7b**.

ρ$_1$ decreases in absolute value up to 60 K, and then continuously increases. Its variation is due to the competition of the $\rho_{PHE}$ and the $\rho_{SMR\parallel}$ components. A nominal longitudinal component ($\rho_{offset}$) has been subtracted in order to supress its variation with temperature.

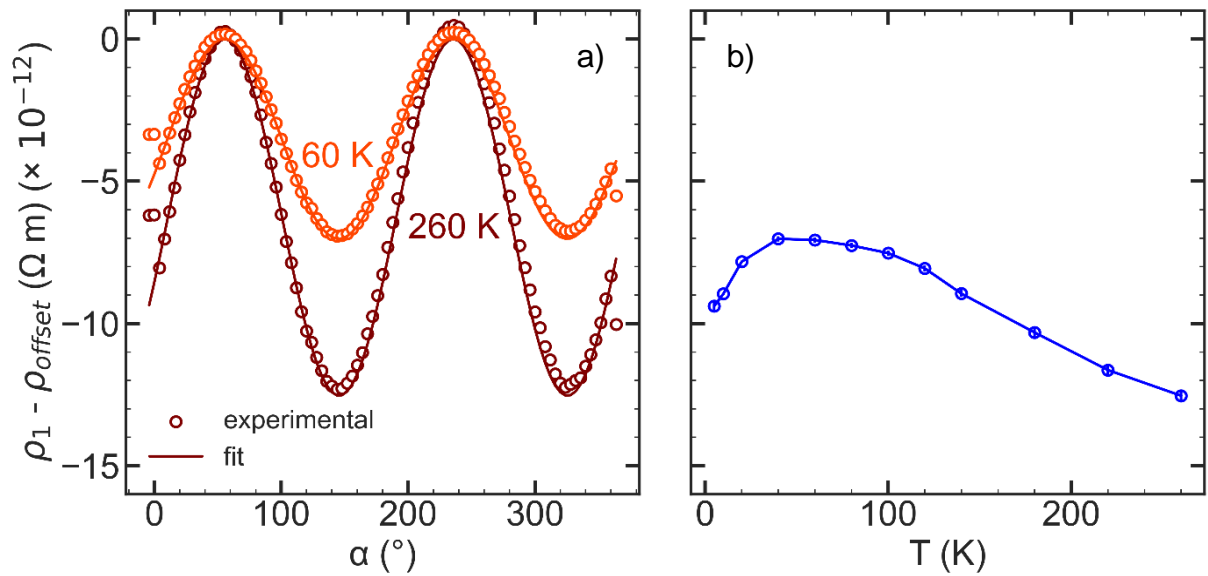

**Figure 7**. a) In-plane angle-dependent magnetoresistance for an applied magnetic field of 7 T at two different temperatures: 60 and 260 K. b) Amplitude of the oscillation vs. T.

## Acknowledgements

This work of the Interdisciplinary Thematic Institute QMat, as part of the ITI 2021 2028 program of the University of Strasbourg, CNRS and Inserm, was supported by IdEx Unistra (ANR 10 IDEX 0002), and by SFRI STRAT'US project (ANR 20 SFRI 0012) and EUR QMAT ANR-17-EURE-0024 under the framework of the French Investments for the Future Program. DP has benefit support from the French National Research Agency (ANR) through the ANR-21-CE08-0021-01 'ANR FOXIES'. The authors acknowledge the collaboration of the 'Meb-Ccro' and 'DRX' platforms of the IPCMS. This work was also supported by the French National Research Agency (ANR) through the ANR-18-CECE24-0008-01 'ANR MISSION' and the No. ANR-19-CE24-0016-01 'Toptronic ANR'. Some devices in the present study were patterned at Institut Jean Lamour's clean room facilities (MiNaLor). These facilities are partially funded by FEDER and Grand Est region through the RANGE project.

## Authors declarations

The authors have no conflicts to disclose.

## Data availability

The data that support the findings of this study are available from the corresponding author upon reasonable request.